\documentclass[10pt, twocolumn, conference, compsoc]{IEEEtran}
\usepackage{indentfirst}
\usepackage{graphics}
\usepackage{epsfig}
\usepackage{url}
\usepackage{array}
\usepackage{setspace}
\usepackage{subfig}
\begin{document}

\title{ScALPEL: A Scalable Adaptive Lightweight Performance Evaluation Library for application performance monitoring}
\author{Hari K. Pyla, Bharath Ramesh, Calvin J. Ribbens and Srinidhi Varadarajan\\
Department of Computer Science, Virginia Tech.\\
\{harip, bramesh, ribbens, srinidhi\}@vt.edu}
\date{February 27, 2009}
\maketitle

\subsection*{\centering Abstract}
{\em
As supercomputers continue to grow in scale and capabilities, it is becoming increasingly difficult to isolate processor and system level causes of performance degradation. Over the last several years, a significant number of performance analysis and monitoring tools have been built/proposed. However, these tools  suffer from several important shortcomings, particularly in distributed environments. In this paper we present ScALPEL, a Scalable Adaptive Lightweight Performance Evaluation Library for application performance monitoring at the functional level. Our approach provides several distinct advantages. First, ScALPEL is portable across a wide variety of architectures, and its ability to selectively monitor functions presents low run-time overhead, enabling its use for large-scale production applications. Second, it is run-time configurable, enabling both dynamic selection of functions to profile as well as events of interest on a per function basis. Third, our approach is transparent in that it requires no source code modifications. Finally, ScALPEL is implemented as a pluggable unit by reusing existing performance monitoring frameworks such as Perfmon and PAPI and extending them to support both sequential and MPI applications.
}
\section{Introduction}

Millions of dollars are spent each year in building faster HPC systems to reduce computation time for a wide range of computational science and engineering applications. However, only a few select benchmarks achieve anything close to peak performance on these high-end resources, with most applications running at a small fraction of the advertised peak speed. Increasing the usable capacity of high-end computational resources necessitates the use of performance measurement and analysis tools that provide detailed sensor data to guide algorithm redesign and optimization. 

Performance analysis of even sequential single node applications is complicated by several factors including cache hierarchy, data placement, resource usage and TLB misses to name a few. This problem is significantly exacerbated for large-scale parallel applications that face variations in scheduling and message transfer latency.  

Simple UNIX commands and basic tools like \emph{time} and gprof~{\footnotesize\cite{gprof_website}} provide preliminary information such as the total amount of time spent in a particular code segment. However, in order to optimize the application this information is not sufficient. The total time reported by these tools is a function of several factors which are often not visible in the source code but happen dynamically when the code is executed on a particular platform. Hardware performance counters provide valuable insights into various performance aspects of applications; they report the ``cause(s)" rather than just the ``effect(s)".

Traditional tools such as Perfmon~{\footnotesize\cite{perfmon2_website}}, PAPI~{\footnotesize\cite{Browne2000,Dongarra2003}}, Perfsuite~{\footnotesize\cite{Kufrin2005}} and many others are commonly used to monitor hardware counters. While such tools are quite useful, they suffer from several shortcomings. 

First, such tools/libraries monitor only a fixed set of events throughout the lifetime a program. Modern x86 processors only allow monitoring of four events at best. These events are hard-coded while profiling the program. Hence, fully exploring the desired event space\footnote{The set of all counters the performance analyst is interested in monitoring.} requires a time consuming iterative process of compiling and running the application many times. Moreover, this process may also require modifications to the source code in order to accommodate a different set of events. Additionally, some of the tools~{\footnotesize\cite{perfmon2_website,Browne2000,Dongarra2003,shark_website,PCL_website}} are not easily usable, since they involve a learning curve to understand the API they provide. 

In addressing these issues, some of the tools~{\footnotesize\cite{perfmon2_website,Kufrin2005}} provide transparency through statistical code sampling and time sharing software multiplexing techniques~{\footnotesize\cite{Azimi2005,May2001,Mathur2005,perfmon2_website}}. While such techniques are useful in exploring the event space, they suffer from tradeoffs involving high accuracy (high sampling rate) and low performance overhead (low sampling rate).  

Most importantly, such techniques are not suitable for a wide range of applications because they report the counter values during a particular phase. This is a serious limitation. Consider a scenario where an application is iterative in nature with varying phases during iterations. Multiplexing events in time may not be synchronized with the phase of the application and hence the true nature of the application is not captured.

Second, existing tools do not facilitate a complete runtime configuration of hardware performance counters. The counters are defined at compile time and cannot be modified during execution of the program. 

Third, most of tools are based on techniques that involve a significant overhead. These techniques commonly involve interrupts, timers, break-points, etc. Such techniques affect the critical path of execution and often result in significant monitoring overhead.

Finally, most of the tools do not provide runtime access to the counters. The raw values obtained from the counters are usually reported after program termination. The lack of such information prevents applications from making any runtime decisions based on performance characteristics. Libraries such as PAPI and Perfmon provide an API to read the counters. However, the counters cannot be accessed asynchronously since the API function calls must be embedded in the code and hence must be specified at compile time.

Even though there are a plethora of performance monitoring tools, analyzing the performance of parallel applications is still a tedious task for two reasons. First, parallel applications are inherently complex. Second, as discussed previously, the tools are not flexible enough and do not simplify the task of performance analysis. In order to simplify this problem and address the above mentioned issues, we present a compiler directed tool called ScALPEL to monitor hardware performance counters. \emph{Our intent is not to develop yet another performance monitoring tool, but instead to develop an approach that can be easily combined with existing tools such as Perfmon and PAPI and make them more usable.}

In this paper, we propose a simple solution that addresses the above problems and makes the following contributions:

\begin{itemize}
\item {\bf Simple:} \emph{We provide an approach that simplifies the process of performance monitoring.}
\item {\bf Lightweight:} \emph{We propose and demonstrate a new compiler driven technique that reduces the performance monitoring overhead significantly, compared to traditional approaches.}
\item {\bf Dynamic:} \emph{We provide a technique that facilitates configuration and access of hardware performance counters at runtime.}
\item {\bf Portable:} \emph{We propose an approach that is portable and transparent and can be seamlessly plugged into existing libraries.}
\end{itemize}

The rest of the paper is organized as follows, we discuss related work in Section~\ref{section_related_work}. We present the details of our approach and implementation in Section~\ref{section_our_approach}. In Section~\ref{section_results} we present our evaluation, case study and experimental results. In Section~\ref{section_limitations} we present the limitations of our approach and finally in Section~\ref{section_conclusion} we summarize our conclusions.

\section {Related Work}\label{section_related_work}
To place our work in the context of existing research and to help clearly understand our contributions, we classify existing approaches based on the level of their implementation (software and hardware).

Software tools can be classified into libraries, tools and simulators. Libraries such as PAPI~{\footnotesize\cite{Browne2000,Dongarra2003}}, Perfmon (\emph{libpfm})~{\footnotesize\cite{perfmon2_website}} and PCL~{\footnotesize\cite{PCL_website}} provide APIs that are well documented and easily accessible. However, they suffer from several important drawbacks. First, they do not offer any transparency to the performance analyst. The performance analyst usually has to go through the tedious task of modifying the program and re-compiling it several times to monitor all the counters. 

On the other hand, open source tools such as Perfmon (Pfmon)~{\footnotesize\cite{perfmon2_website}}, Apple's Shark~{\footnotesize\cite{shark_website}}, Perfsuite~{\footnotesize\cite{Kufrin2005}}, ProfileMe~{\footnotesize\cite{Dean1997}}, TAU ~{\footnotesize\cite{Shende2006}} and commercial tools such as Intel's VTune~{\footnotesize\cite{intelvtune_website}} are usable, transparent and provide elaborate graphical results. However they often suffer from significant profiling performance overhead. Most importantly most of these tools install break points in the application and hence are not suitable for profiling recursive or nested function calls. Our experiments with Perfmon indicate that such techniques can lead to significant performance overheads. Another approach proposed by Zagha et al.~{\footnotesize\cite{Zagha1996}} provides minimal information at a per process level. Such implementations have lower performance overhead at the expense of granularity and portability. Most importantly, Zagha's implementation is specific to the MIPS architecture and it is not portable across other commonly used architectures such as x86. We take a completely different approach by providing a much more fine-grained, portable and architecture independent solution.

Software tools can also be classified based on the underlying profiling technique. Several approaches~{\footnotesize\cite{Azimi2005, Kim2007,perfmon2_website,shark_website,Shende2006}} are based on statistical sampling. Such implementations are often faced with a choice between accuracy and performance penalty. Our approach does not involve sampling; in fact, we leverage support from the compiler to profile the application. 

Other implementations of performance tools rely on simulators~{\footnotesize\cite{Mellor2001, Lebeck1994, Rosenblum1997, Rose1998}}. They provide limited information with a limited set of events for the modeled processors. Such an approach is often slow and not scalable across different nodes in a cluster. We choose to implement our technique using real systems running on commonly used architectures; moreover, our approach is easily scalable across multiple cores and nodes in a cluster. 

Finally, other approaches~{\footnotesize\cite{Noordergraaf2002, Dean1997}} include special purpose hardware to monitor the counters. Such techniques incur relatively less performance overhead compared to software based approaches. However, they are not usable in that they require special hardware, which is expensive and usually not a feasible solution to install on all the nodes in a cluster. Furthermore, they require architecture-specific software to use their hardware. In contrast, our approach is software based and can run as a module on existing tools and libraries. 

\section {Our approach}\label{section_our_approach}
In this section we discuss the design and implementation details of ScALPEL. We also explain the rationale behind some of our design choices.

\subsection {Overview}

\begin{figure*}[!ht]
\centering
\subfloat[overview]{\label{fig:overview}\includegraphics[scale=0.45]{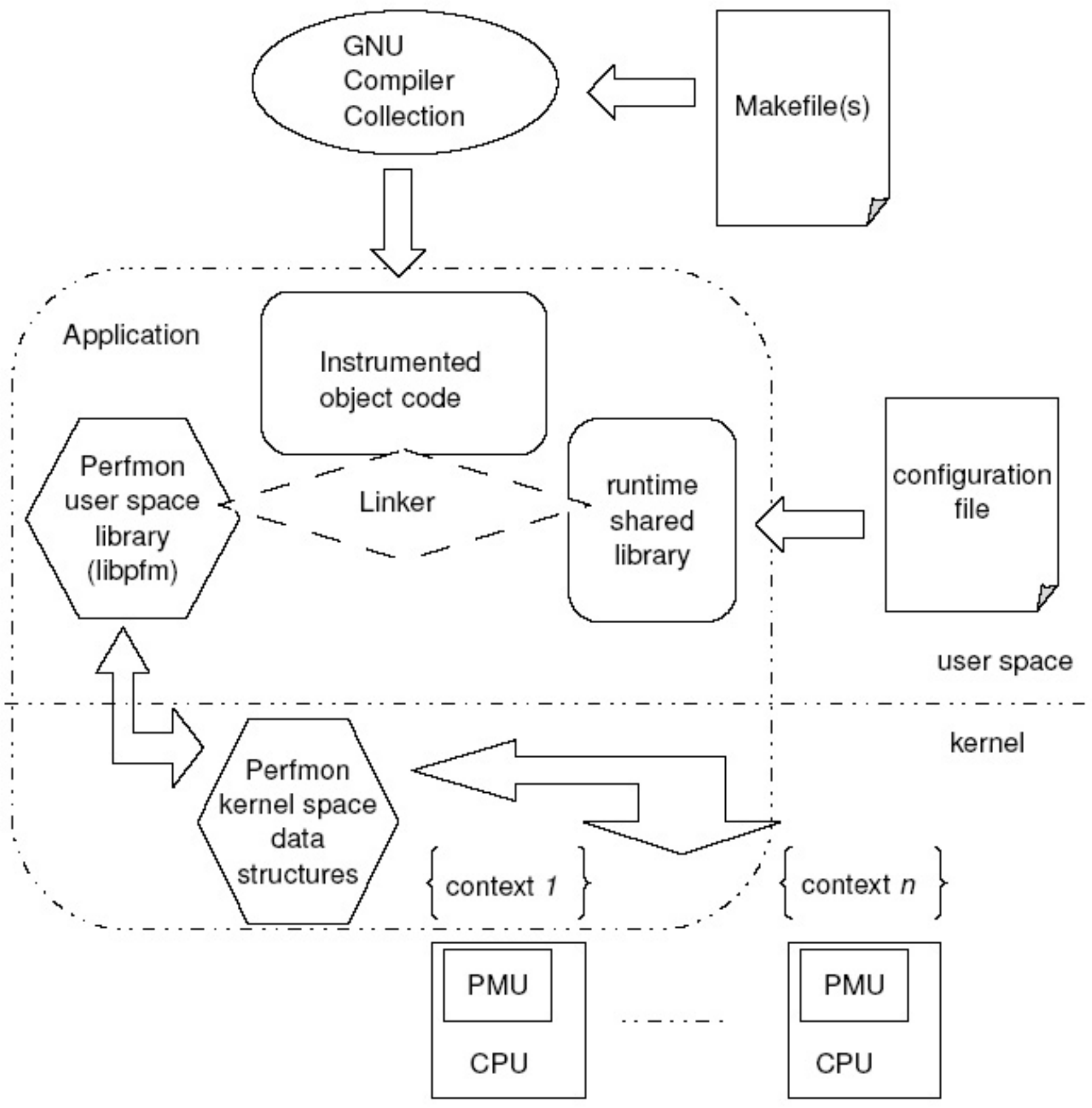}}
\subfloat[control flow]{\label{fig:control_flow}\includegraphics[scale=0.3]{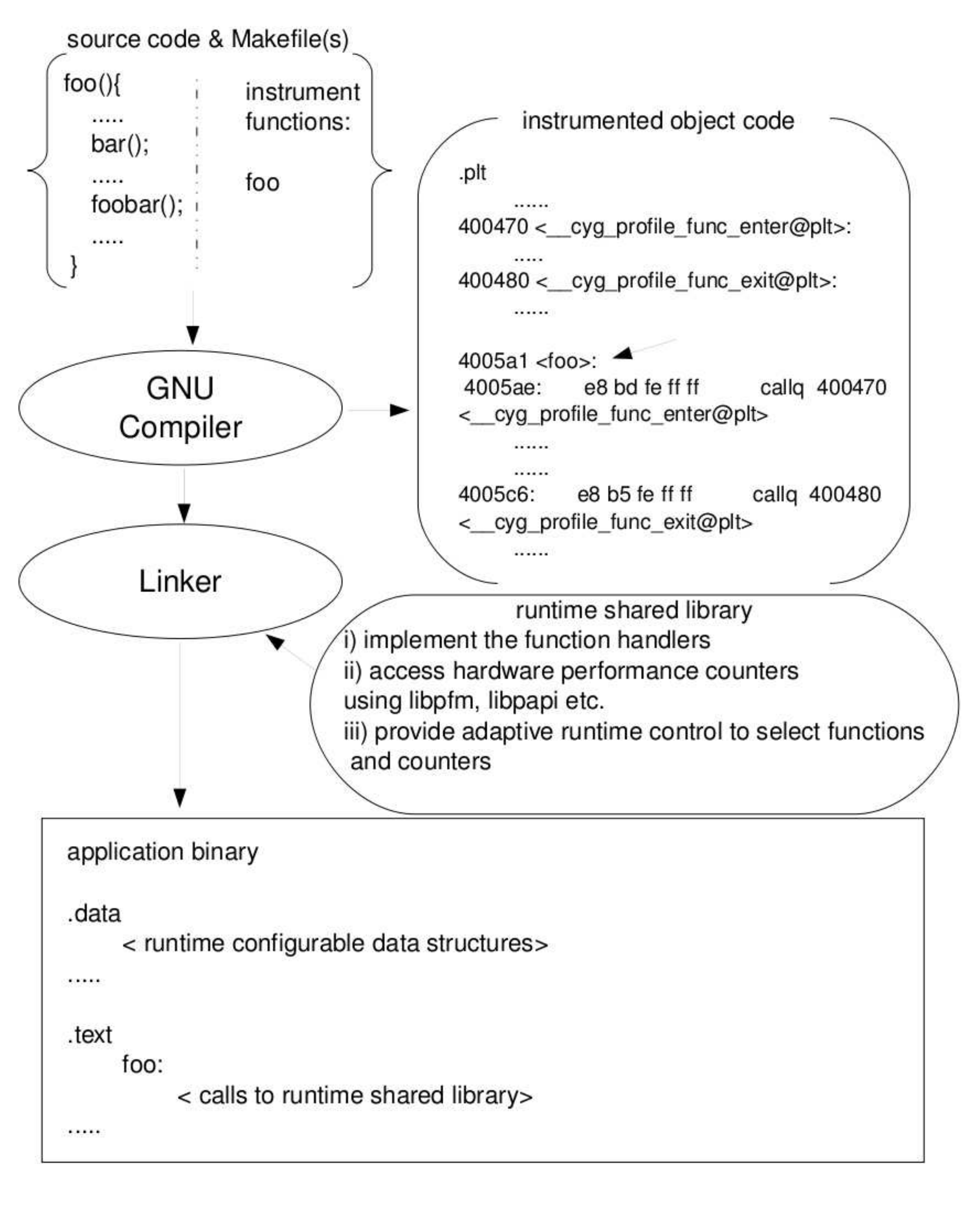}}
\caption{Overview and instrumentation process of ScALPEL.}
\end{figure*}

We present an overview of our approach in Figure~\ref{fig:overview}. Our design includes support for instrumenting source code and adaptively configuring hardware performance counters and functions at runtime. We provide a runtime library that implements the instrumentation callbacks and facilitates runtime reconfiguration. We link the application's object code and the runtime library with a pre-existing user-space hardware performance monitoring library to generate the application binary. 

Unlike traditional debugging techniques such as breakpoints, interrupts and timers, we use a compile time technique to instrument source code. We choose this approach because of its low overhead and ease of installing callbacks in the object code. From a performance analyst's perspective, while our current implementation requires source recompilation, it does not require any code modifications. The instrumentation occurs at the object code level. We designed our approach to be generic in that it can be extended to other existing compilers and performance monitoring utilities. We choose to use the GNU compiler collection and Perfmon for several reasons. First, they are both open source and are commonly available. While Intel Compilers~{\footnotesize\cite{intel_for_compiler_website}} also support function level instrumentation, they are commercial products and not freely available. Second, the design and interface for the Perfmon library is well written and easy to understand. Finally, extending our approach to include other libraries such as PAPI requires little or no effort. 

We present a detailed discussion of ScALPEL's instrumentation methodology and its runtime system in the following subsections.

\subsection {Compiler directed instrumentation}
The GNU compiler~{\footnotesize\cite{gnu_compiler_collection_website}} supports function level instrumentation through its code generation options~{\footnotesize\cite{gcc_opt_website}}. Based on these options, the compiler instruments either all or selected~{\footnote{selective functional profiling is available in GCC 4.3.2 onwards}} functions in an application. In the case of selective functional profiling, the performance analyst is required to identify the functions that are of interest and enumerate the rest of the functions as arguments to the compiler command-line. The compiler then instruments these functions with function handlers or callbacks upon function entry and exit. These function handlers are embedded in the object code. For example, as shown in Figure~\ref{fig:control_flow}, function \emph{foo} is instrumented by the compiler by inserting callbacks immediately upon entry and just before exit. This concludes the first step where the application is aware of functions that can be monitored.

We provide two degrees of freedom to the user: the ability to dynamically change the ``events" and the ``functions" to monitor at runtime. It is important to note that the list of functions specified at compile time defines the set of all functions that are intercepted. Our design provides enough flexibility to let the user select a subset from this set of functions at runtime. If the set cannot be determined at compile time then an alternative is to choose all functions to be intercepted. This results in additional but minor overhead in most cases. However, the actual overhead is strictly application dependant. We present a detailed discussion of the performance implications of these alternatives in the Section~\ref{section_results}.

We define a \emph{context} for each function that the user is interested in monitoring. In ScALPEL a context is centered around a function. The context includes the function name, total number of events, the events, number of subevents for a given event and its corresponding subevents. In Table 1 we present a semantic representation of a typical context. This contextual information is supplied by the user in a configuration file and can be altered at anytime during the application's execution. 

\begin{table}[h]
\caption{\small Layout of a sample configuration file}
\centering
\begin{tabular}{ p{3.0in} } 
\hline\hline
\small\begin{verbatim}
BINARY=my_a.out      // name of the binary
NO_FUNCTIONS=1       // number of functions 

[FUNCTION]
FUNC_NAME=foo        // name of the function	

NO_EVENTS=2          // total number of events

[EVENT]	
ID=DATA_CACHE_MISSES // the event name or id
NO_SUBEVENTS=0       // number of subevents
[/EVENT]

[EVENT]              // begin event information
ID=DISPATCHED_FPU
NO_SUBEVENTS=3
[SUBEVENT]           // list of subevents
ID=OPS_ADD
ID=OPS_ADD_PIPE_LOAD_OPS
ID=OPS_MULTIPLY_PIPE_LOAD_OPS
[/SUBEVENT]
[/EVENT]             // end of event

[/FUNCTION]          // end of function

\end{verbatim}\\
\hline
\end{tabular}
\end{table}
\subsection {Runtime library}\label{section_runtime_library}

In this section we discuss the next step in performance monitoring, i.e., how ScALPEL actually monitors the hardware counters and how the context information can be changed at runtime. Recall that during compile time, only function handlers are placed in the object code. The ScALPEL's runtime shared library implements these function entry and exit handlers. When it is first loaded the runtime library creates a list of all function contexts described by the user from the initial configuration file. 

At runtime, for every function in the set, on entry ScALPEL checks if a context corresponding to this function is specified by the user. If a context does not exist then the function continues executing normally. If a context does exist, then it is loaded and monitoring is initiated for the function. ScALPEL retains the context across any recursive or immediate successive calls to the same function. This helps in reducing the monitoring overhead during recursion and also in situations where a function is called several times repeatedly within a loop. ScALPEL stops monitoring the counters on exit of a function. ScALPEL's runtime library uses the user-space Perfmon library (\emph{libpfm}) to monitor performance counters. The hardware counters are specific to the executing process and not for the entire system. In fact they specifically measure the counters during the scope of a function's execution. The callbacks installed during the instrumentation process identify functions by only their addresses. ScALPEL resolves the addresses with function names by reading the symbol table in the object file to present meaningful results to the user. The results include the name of the function, set of events and their corresponding counter values. 

The runtime library is flexible in that it allows the performance analyst to modify the contexts at runtime in several ways. For instance, a new configuration file may be loaded at any time by sending a signal (SIGUSR1) to the application. At this point the runtime system dumps the previous contexts and creates a new set of contexts. New functions can be added or deleted from the list as long as they are from the set specified at compile time. If new functions are added, and whenever they are encountered during execution thereafter, they are monitored. Functions may also be deleted which stops their monitoring. In addition to dynamically configuring the functions and events, ScALPEL also supports multiplexing of events within a function based the number of function calls. 

The runtime library contains other data structures that store the counter information and provide easy access to this data at runtime. This allows the performance analyst to make runtime decisions. By default, the values of hardware counters observed during program execution are written to \emph{stdout} upon termination of the program. In our present implementation we choose to use \emph{libpfm} to monitor the hardware counters; however, other libraries such as \emph{libpapi} can be used with minor code modifications to the runtime system. 

\section {Experimental analysis and results}\label{section_results}
In this section we explain in detail our experimental methodology and the results of our approach. We tested the ScALPEL prototype on the System-G~{\footnotesize\cite{system_g_website}} supercomputer running Linux 2.6.27.10 x86\_64 kernel with the Perfmon kernel patch~{\footnotesize\cite{perfmon_kernel_patch_website}}. Each node in System G has two quad core Intel Xeon processors running at 2.8 GHz and 8GB of main memory. The nodes are interconnected over a QDR Infiniband switching fabric.

We initially present a performance analysis of our approach compared to Perfmon. We then present a case study to show the flexibility and ease of use of ScALPEL. We used the NAS-Parallel benchmark suite~{\footnotesize\cite{nas_benchmark_website}} for performance analysis and LINPACK~{\footnotesize\cite{linpack_website}} as a case study to help illustrate the adaptivity provided by ScALPEL. 

\begin{figure*}[!ht]
\centering
\subfloat[BT (log-lin plot)]{\label{fig:bt:binvcrhs}\includegraphics[scale=0.45]{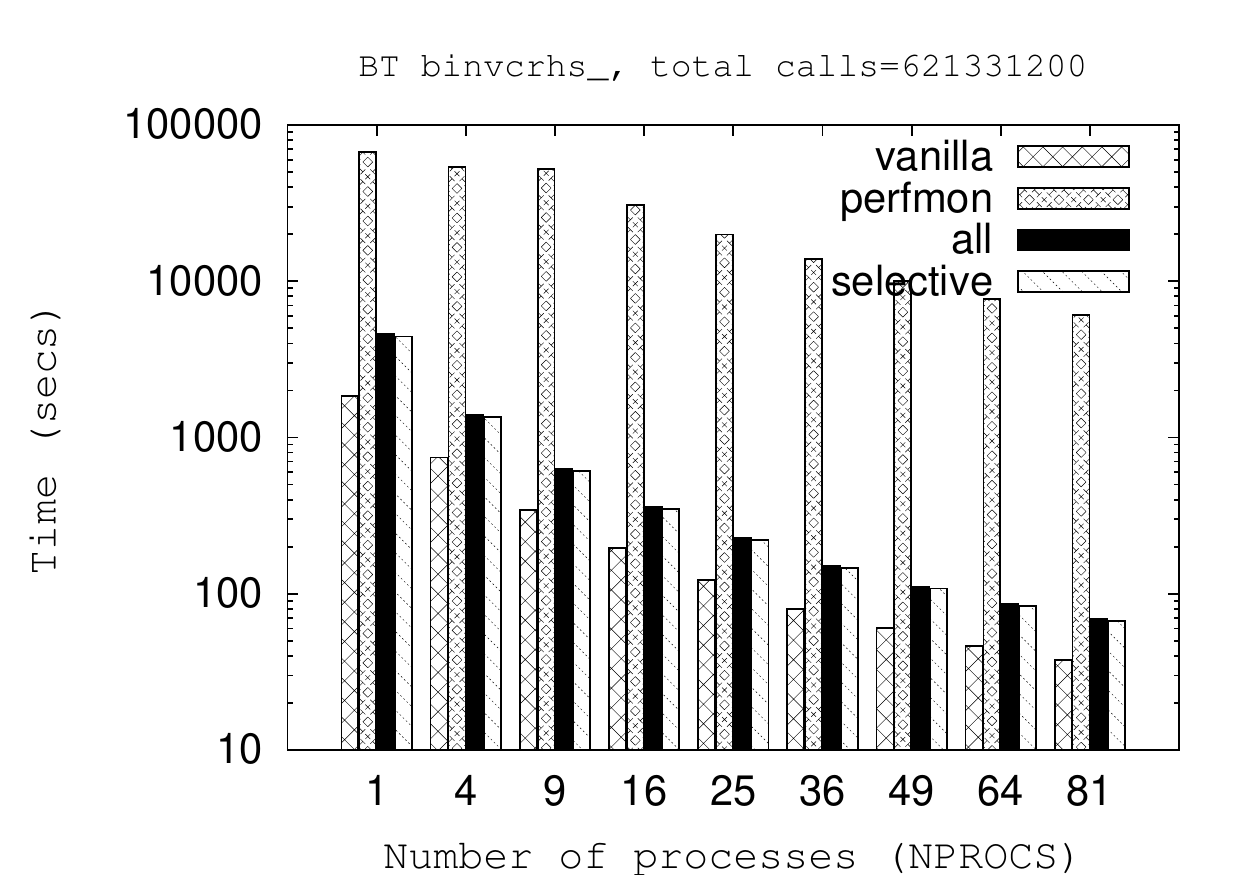}}
\subfloat[IS (log-lin plot)]{\label{fig:is:randlc}\includegraphics[scale=0.45]{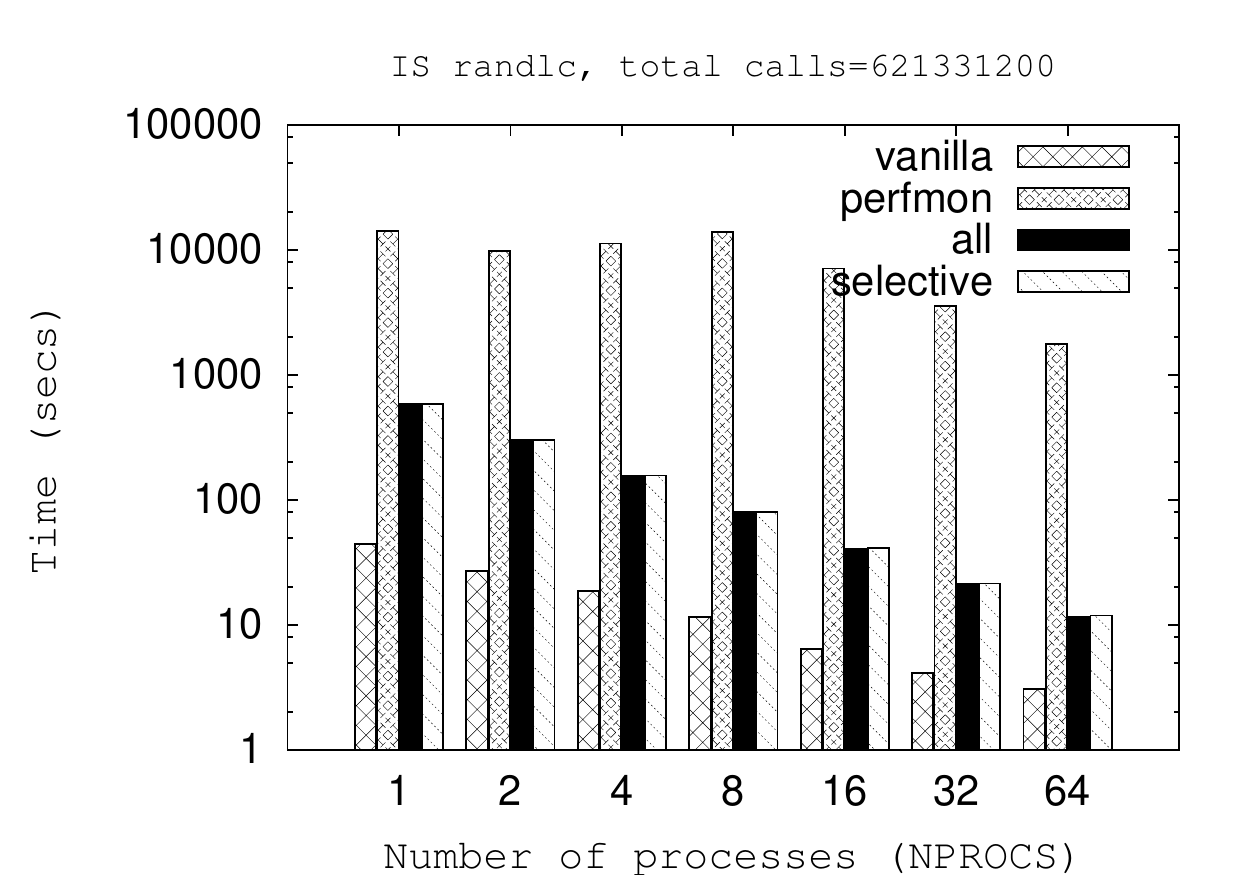}}
\subfloat[FT]{\label{fig:ft:fftz2}\includegraphics[scale=0.45]{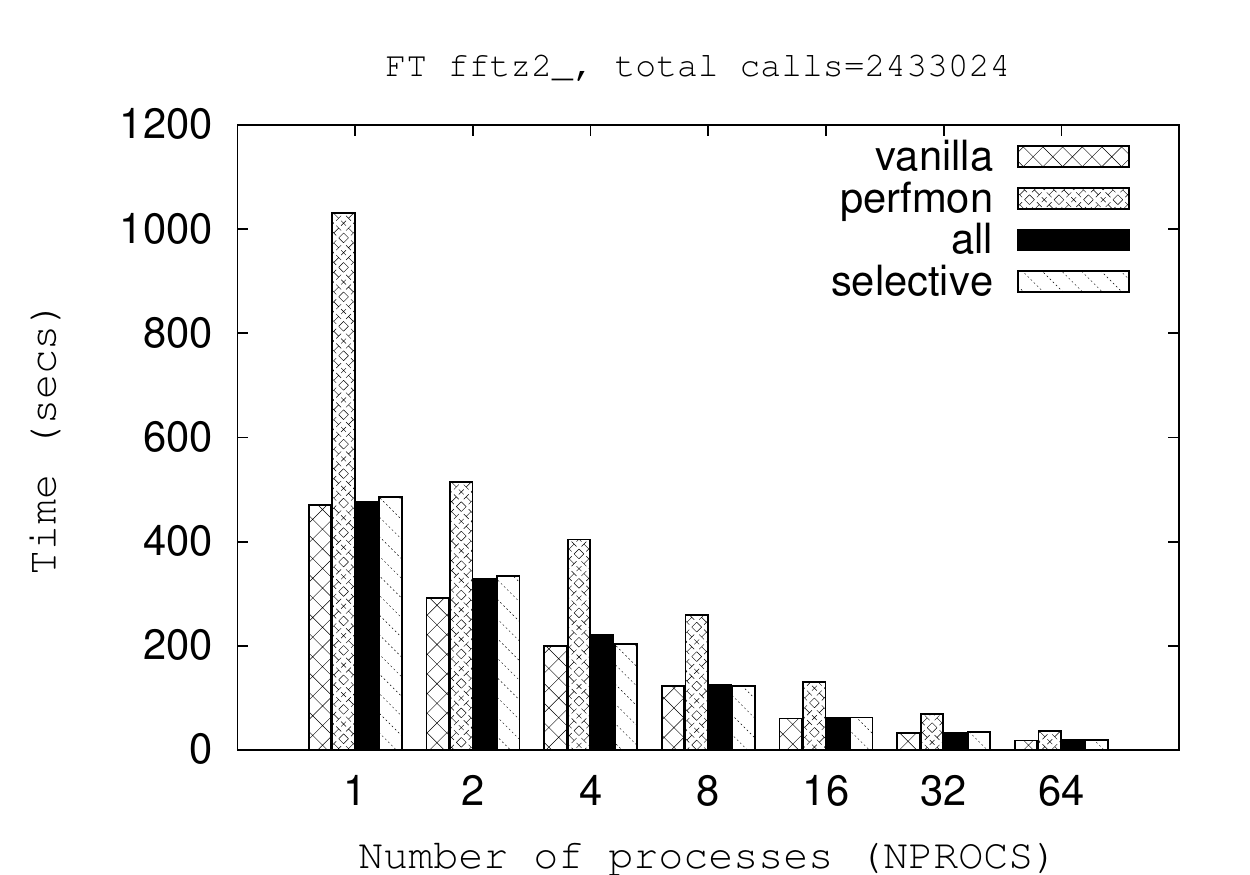}} \\

\subfloat[LU]{\label{fig:lu:exact}\includegraphics[scale=0.45]{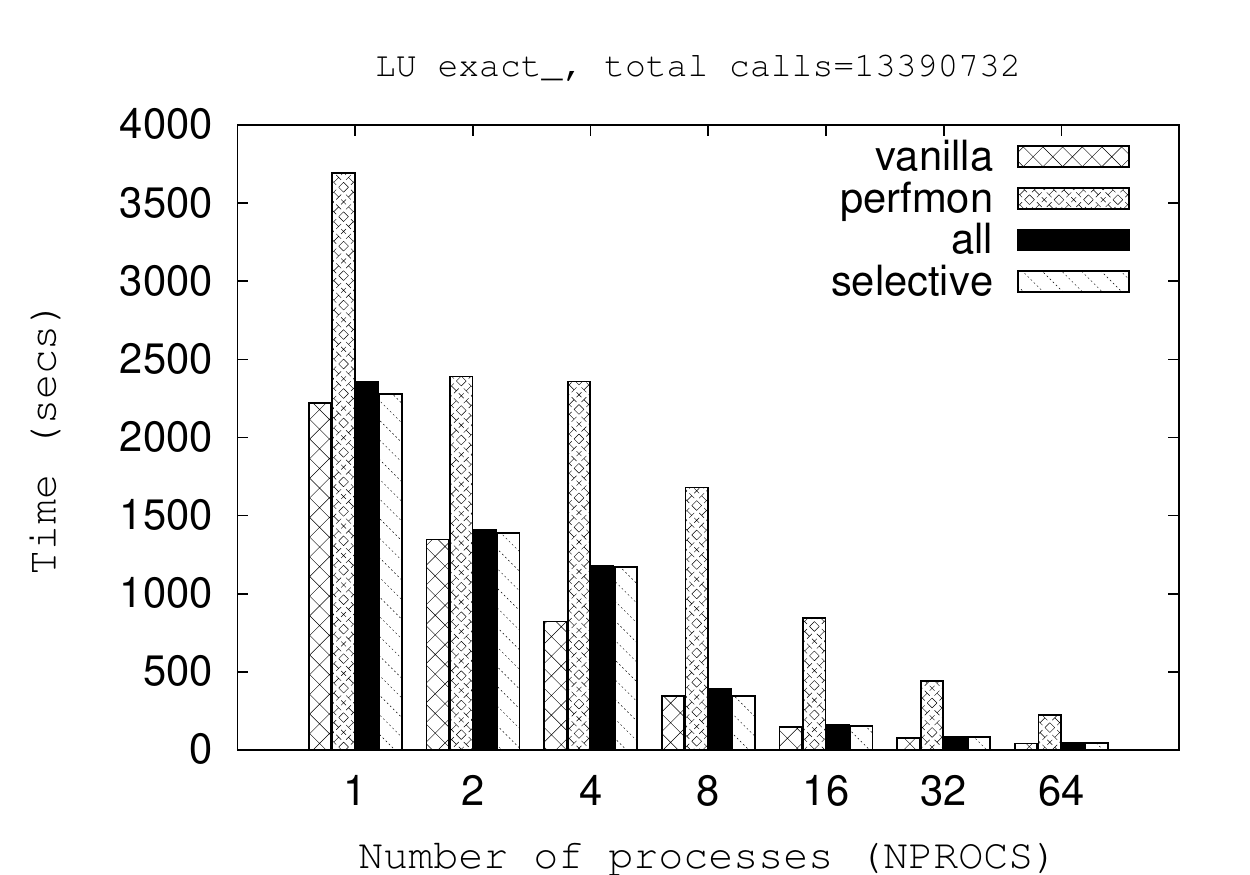}}
\subfloat[MG]{\label{fig:mg:randlc}\includegraphics[scale=0.45]{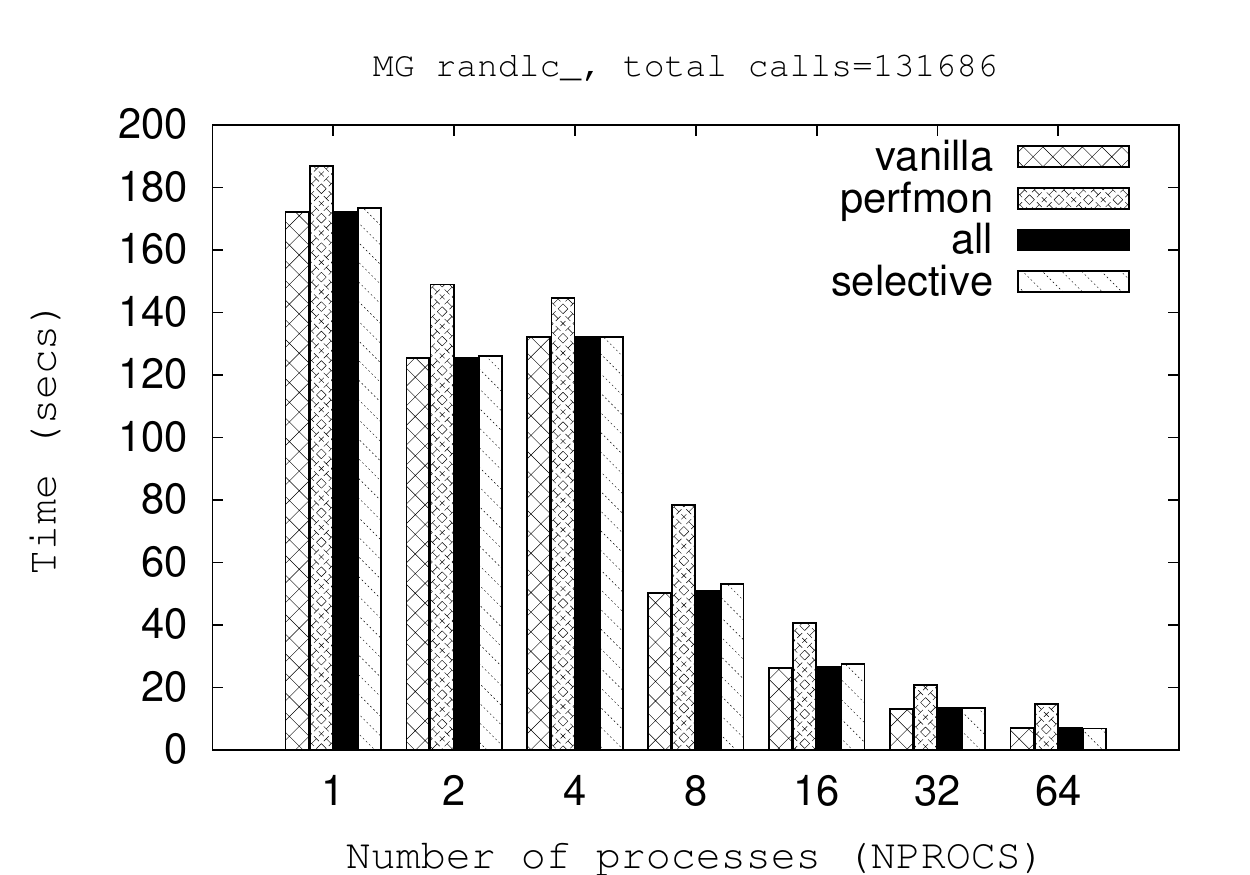}}
\subfloat[SP]{\label{fig:sp:exact_solution}\includegraphics[scale=0.45]{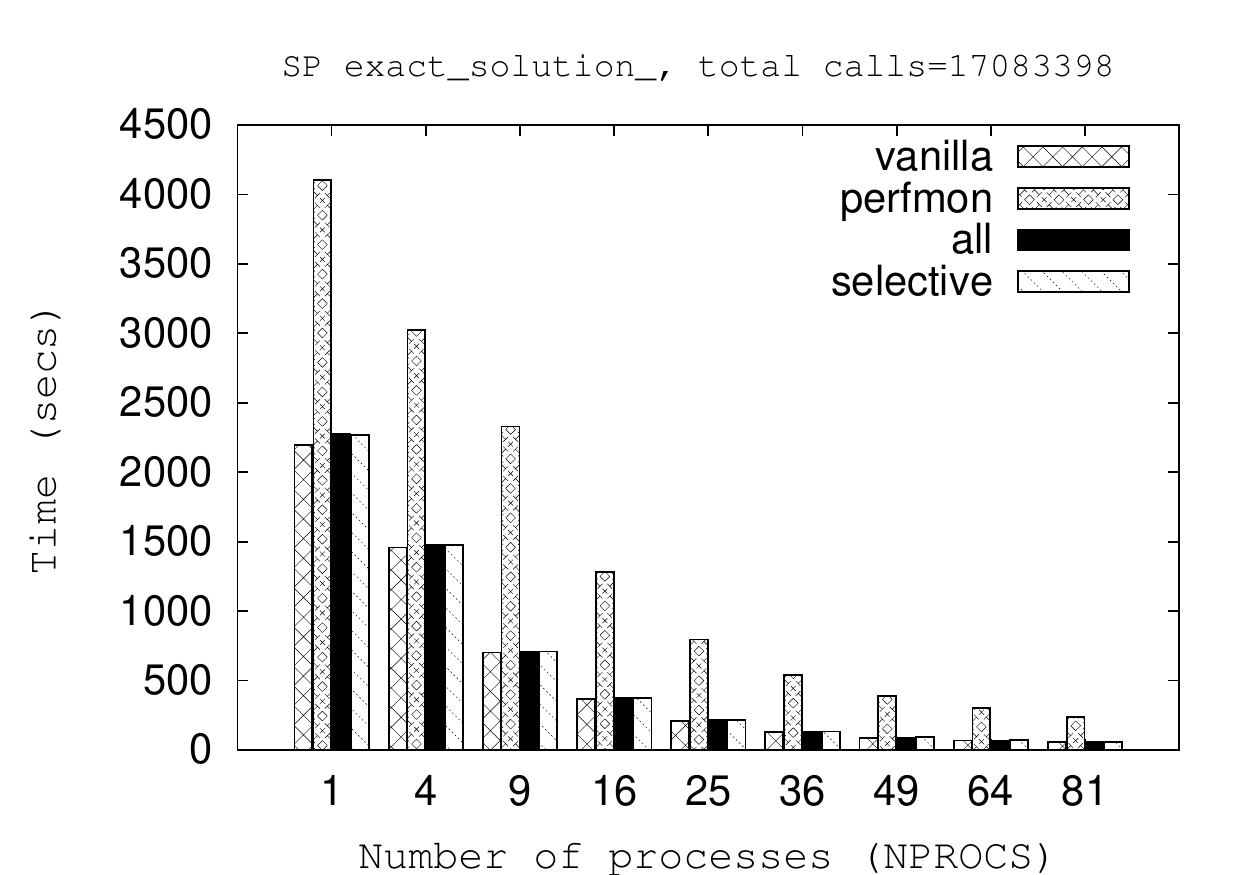}} \\

\caption{Illustrates the performance overhead of various test cases \emph{vanilla}, \emph{perfmon}, \emph{all} and \emph{selective} along with the selected function and its total number of calls. We choose to use $log_{10}$ scale for BT and IS because the overhead of perfmon was too dramatic and the rest of the cases were negligible in the graph otherwise.}
\label{figure2}
\end{figure*}

\subsection{Performance evaluation}
We quantify the performance overhead by measuring the total real time for each execution using the UNIX \emph{time} command. We define the following four test cases;
\begin{itemize}
\item{\emph{vanilla:} does not involve any hardware performance monitoring. The application is run natively}
\item{\emph{perfmon:} measures the hardware counters using Perfmon tool}
\item{\emph{all:} intercepts all the functions in an application, i.e., set of all intercepted but not necessarily monitored functions}
\item{\emph{selective:} intercepts and monitors only a subset of the total functions in an application}
\end{itemize}

In order to fully understand the extent of performance overhead involved in providing complete flexibility to choose both functions and counters, we sub-divide our approach into two categories (\emph{all} and \emph{selective}), depending on the size of the set\footnote{Recall that the size of the set indicates the number of functions chosen for profiling at compile time.}. The \emph{all} and \emph{selective} test cases describe the worst and best case scenarios respectively of our instrumentation overhead. However, it is important to note that they are not necessarily indicative of the performance counter monitoring overhead. To evaluate the monitoring overhead, we choose to monitor a single function at a time throughout the execution of an application. We choose to do this because it is not possible to obtain the hardware performance counter values for individual functions at a per-function level using \emph{perfmon} without involving sampling. Hence, in the case of \emph{all}, we intercept all functions but monitor only one function, thereby paying the cost of monitoring without any gains from the ability to dynamically change the set of monitored functions. Similarly, in the case of \emph{selective} and \emph{perfmon} we intercept and monitor only one function. We monitored the same set of events and same functions across all four test cases.

\begin{figure*}[htp]
\subfloat[BT]{\label{fig:bt:exact_s}\includegraphics[scale=0.45]{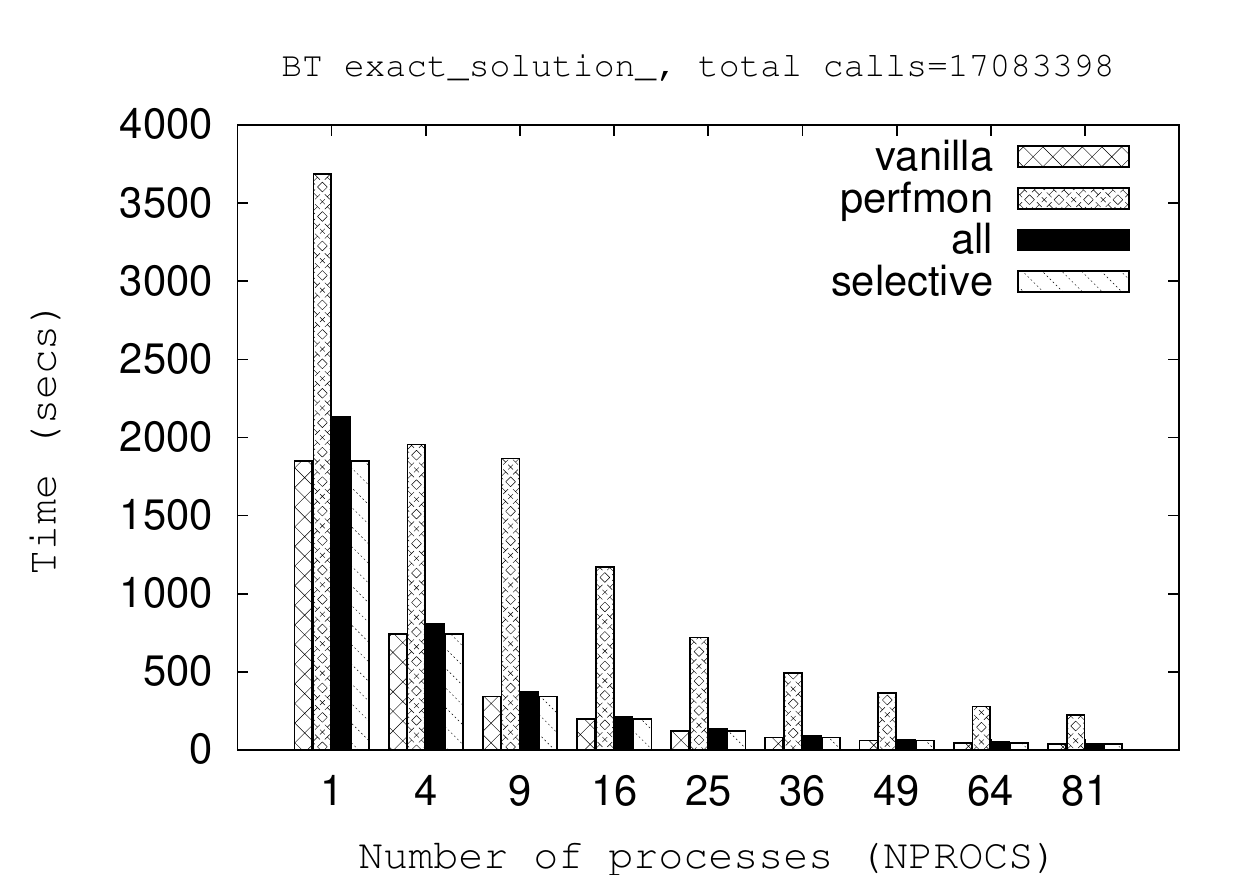}}
\subfloat[BT]{\label{fig:bt2}\includegraphics[scale=0.45]{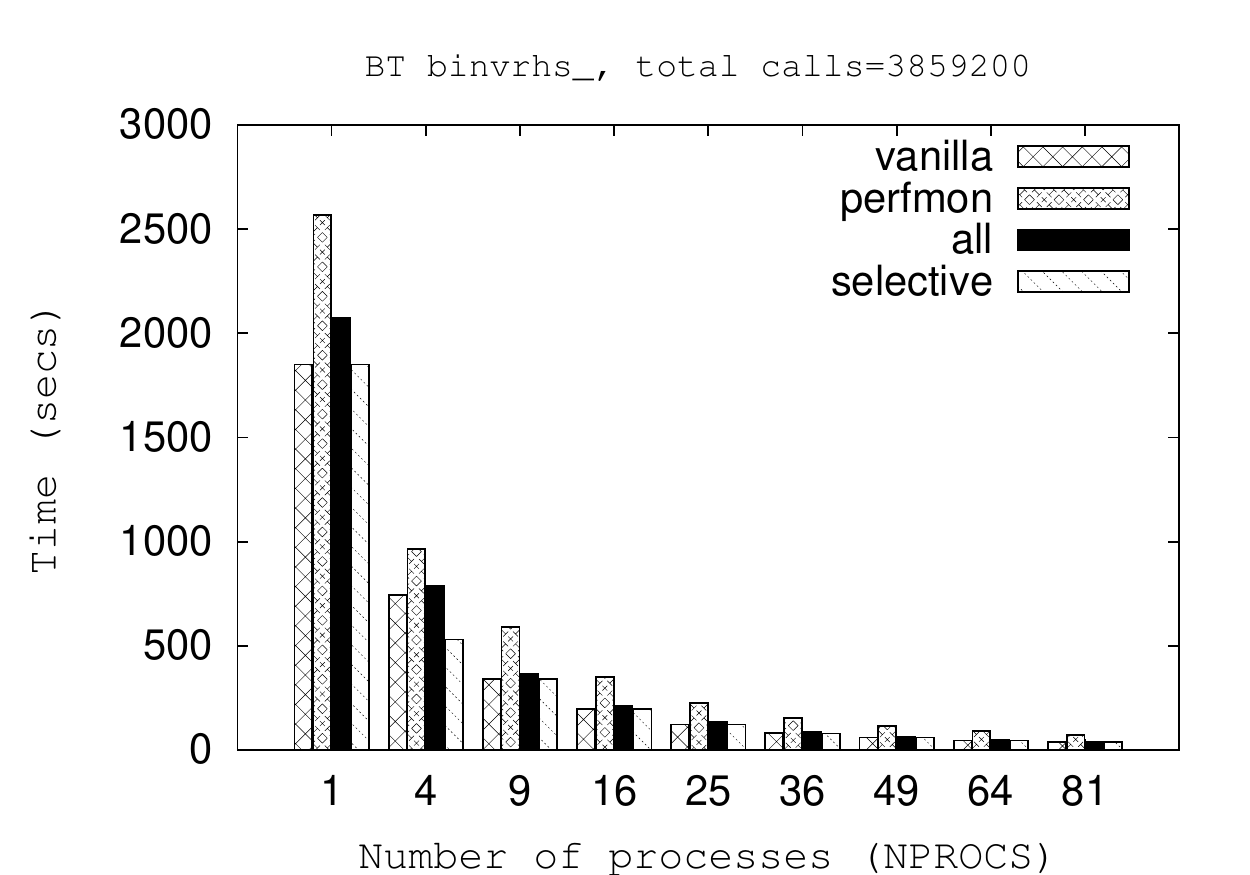}}
\subfloat[CG]{\label{fig:cg2}\includegraphics[scale=0.45]{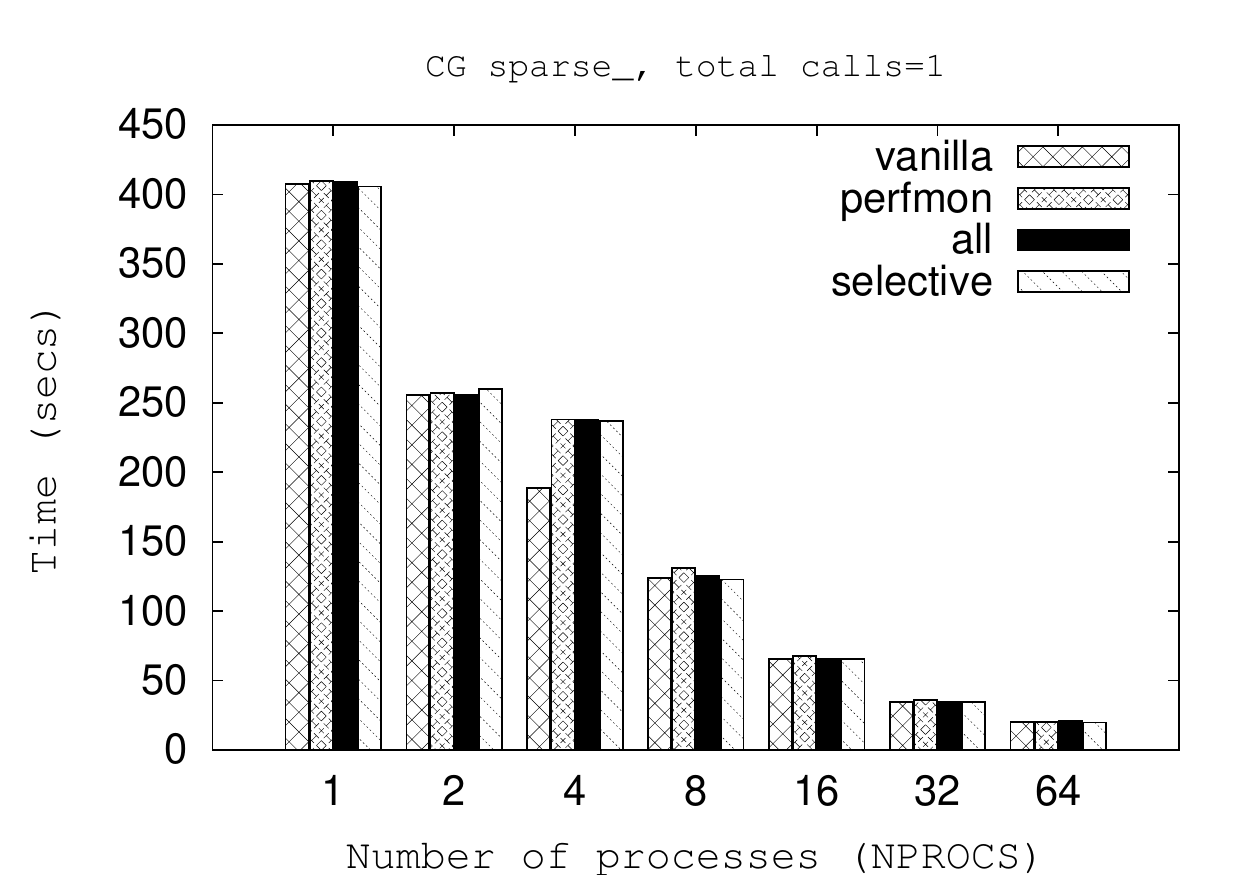}} \\

\subfloat[EP]{\label{fig:ep1}\includegraphics[scale=0.45]{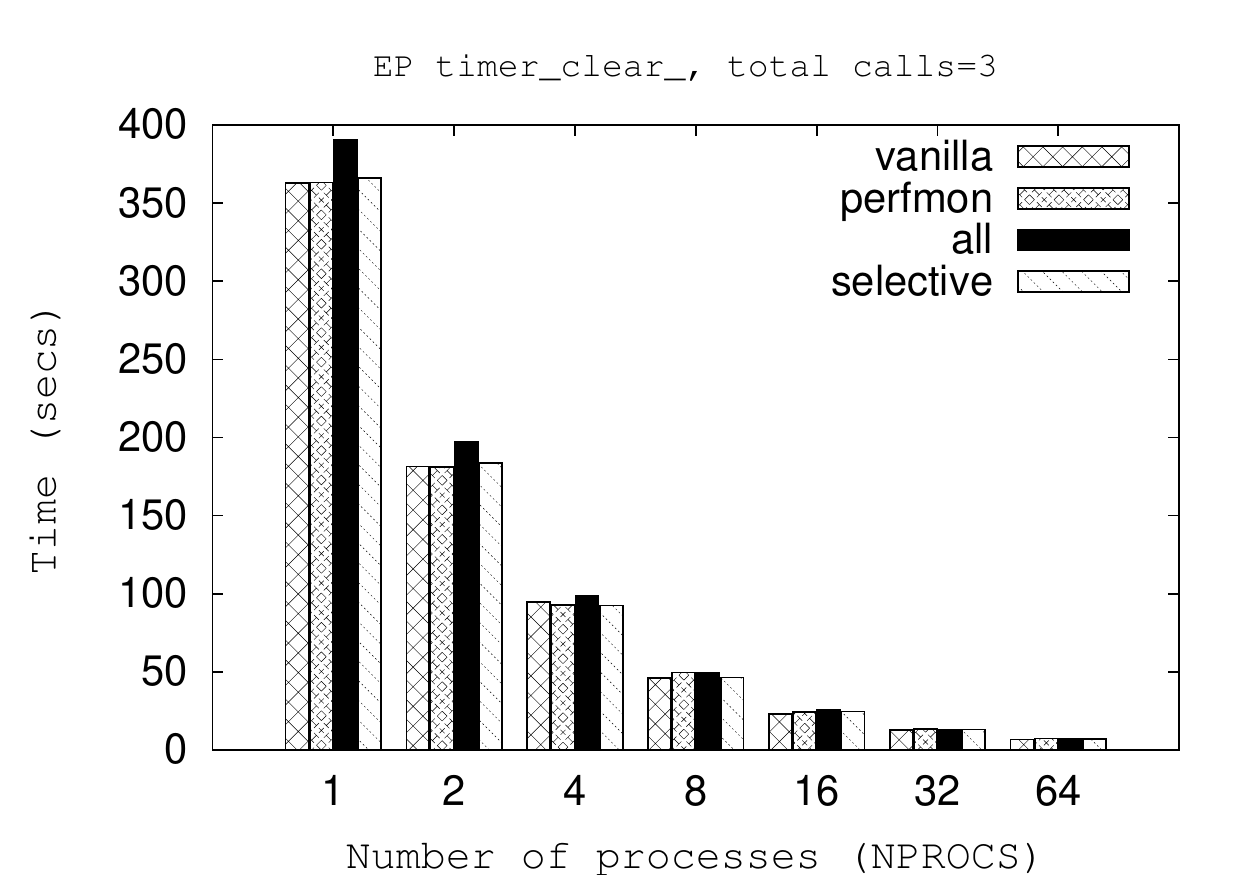}}
\subfloat[EP]{\label{fig:ep2}\includegraphics[scale=0.45]{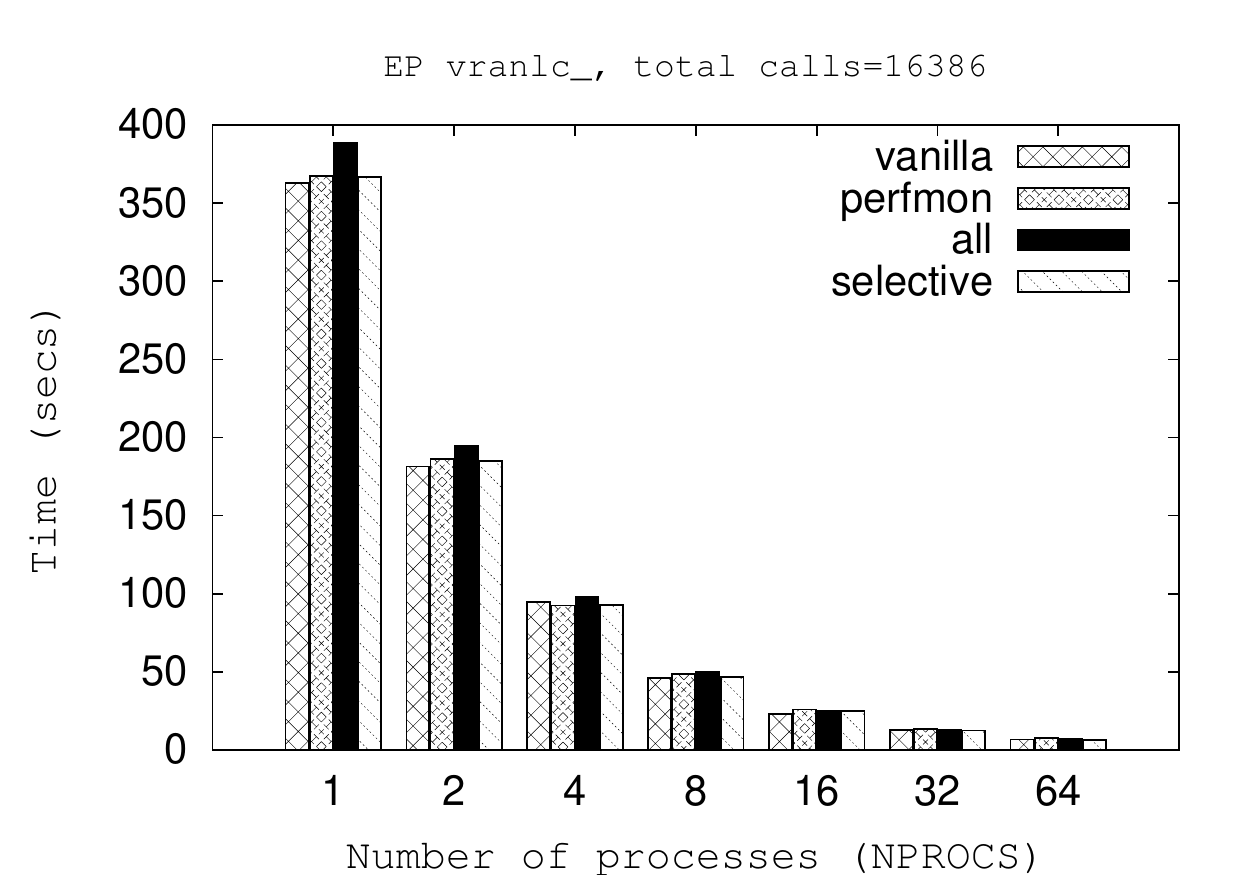}}
\subfloat[FT]{\label{fig:ft2}\includegraphics[scale=0.45]{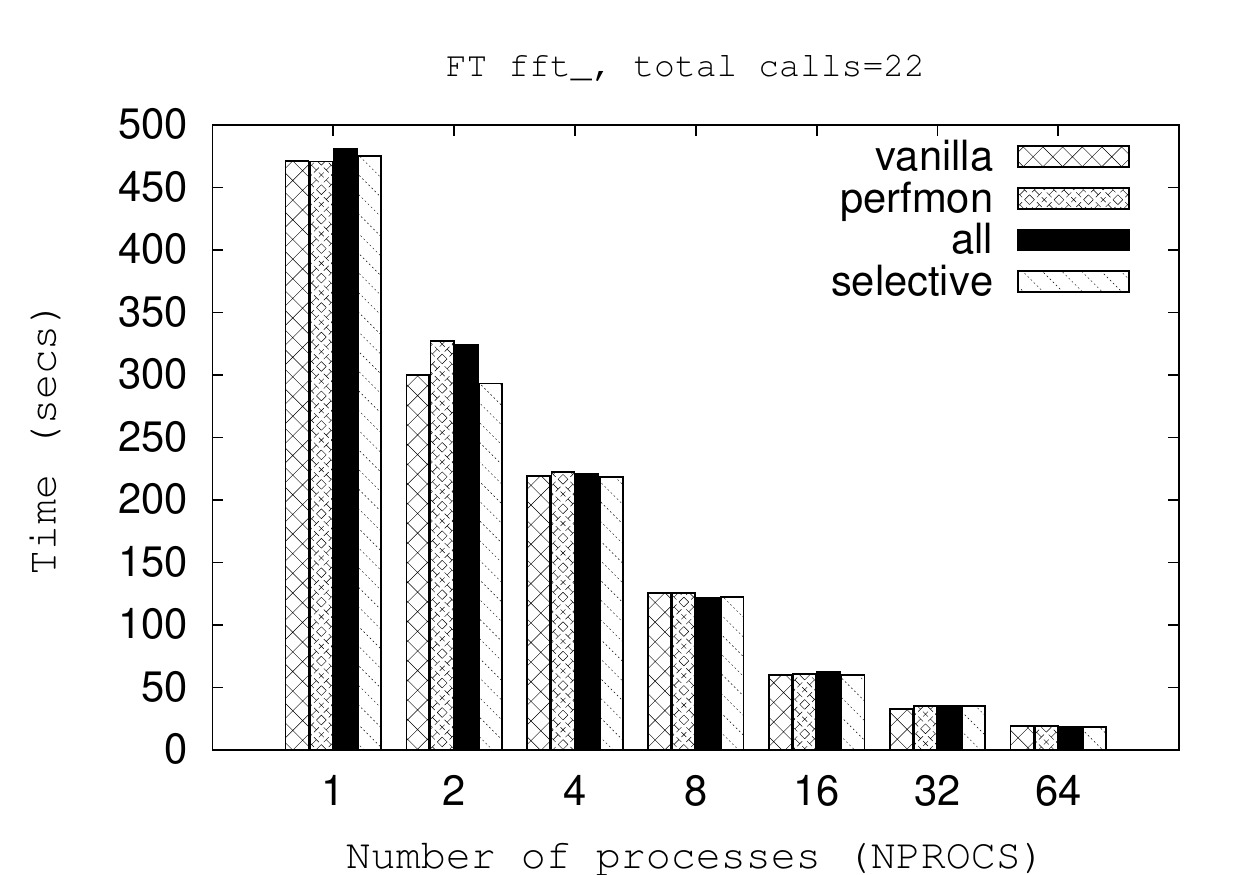}} \\

\subfloat[IS]{\label{fig:is1}\includegraphics[scale=0.45]{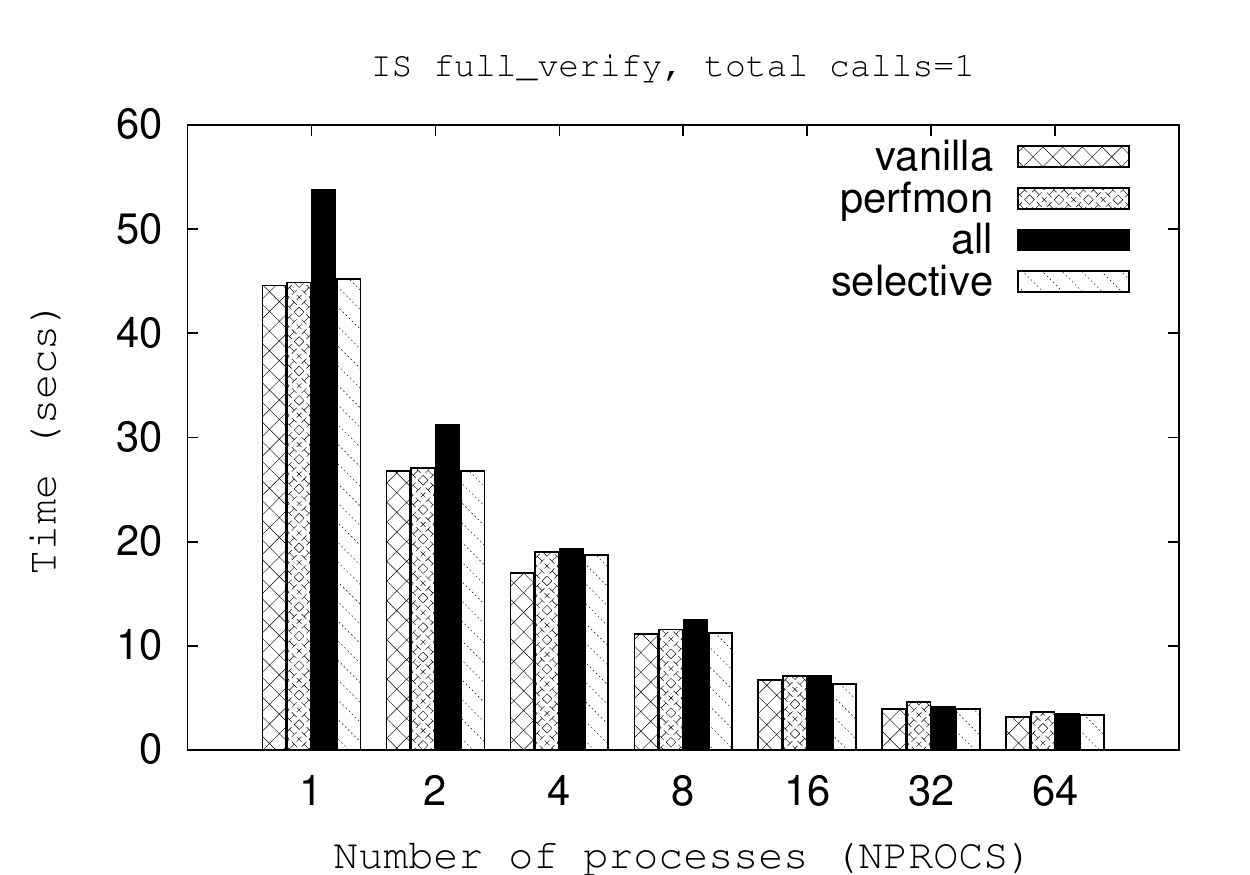}}
\subfloat[IS]{\label{fig:is2}\includegraphics[scale=0.45]{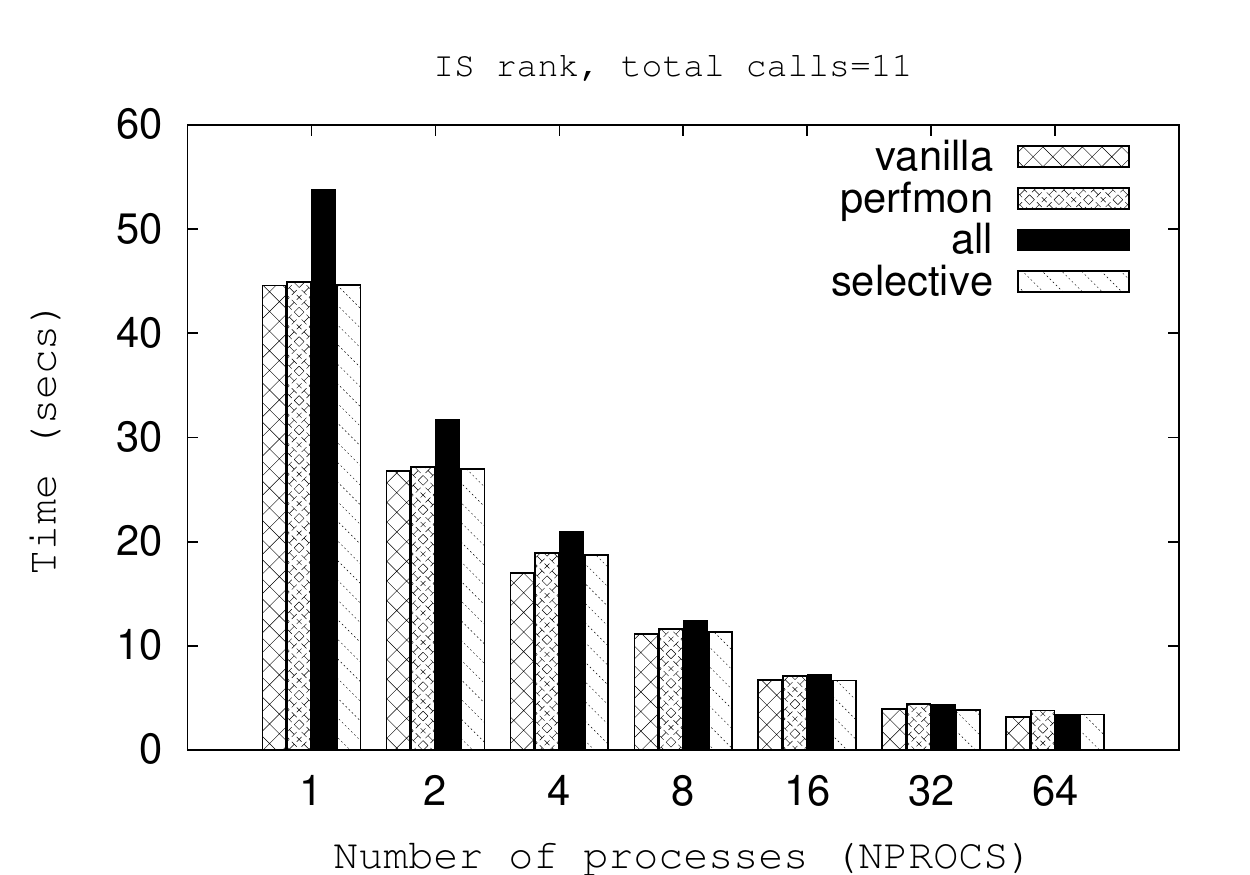}}
\subfloat[LU]{\label{fig:lu2}\includegraphics[scale=0.45]{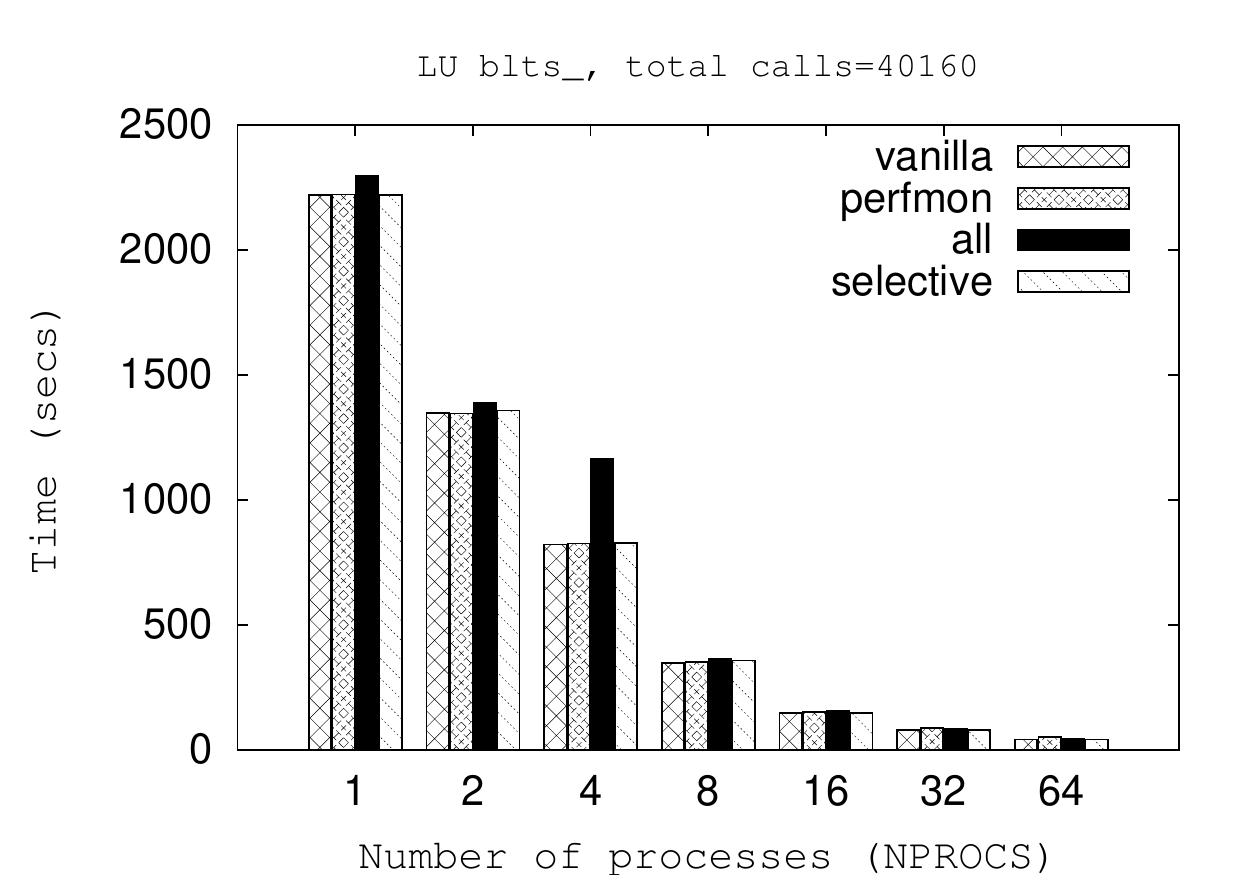}}\\

\subfloat[MG]{\label{fig:mg1}\includegraphics[scale=0.45]{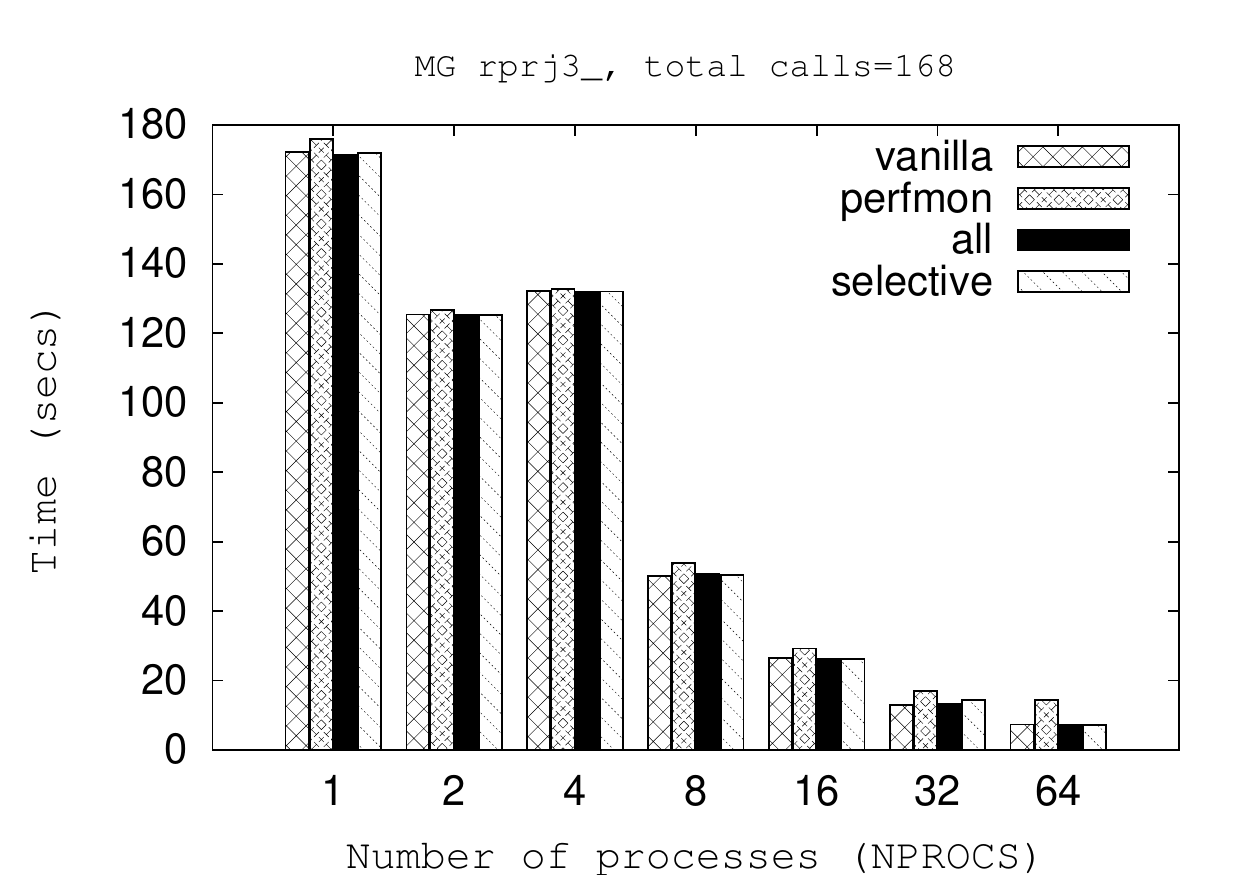}}
\subfloat[MG]{\label{fig:mg2}\includegraphics[scale=0.45]{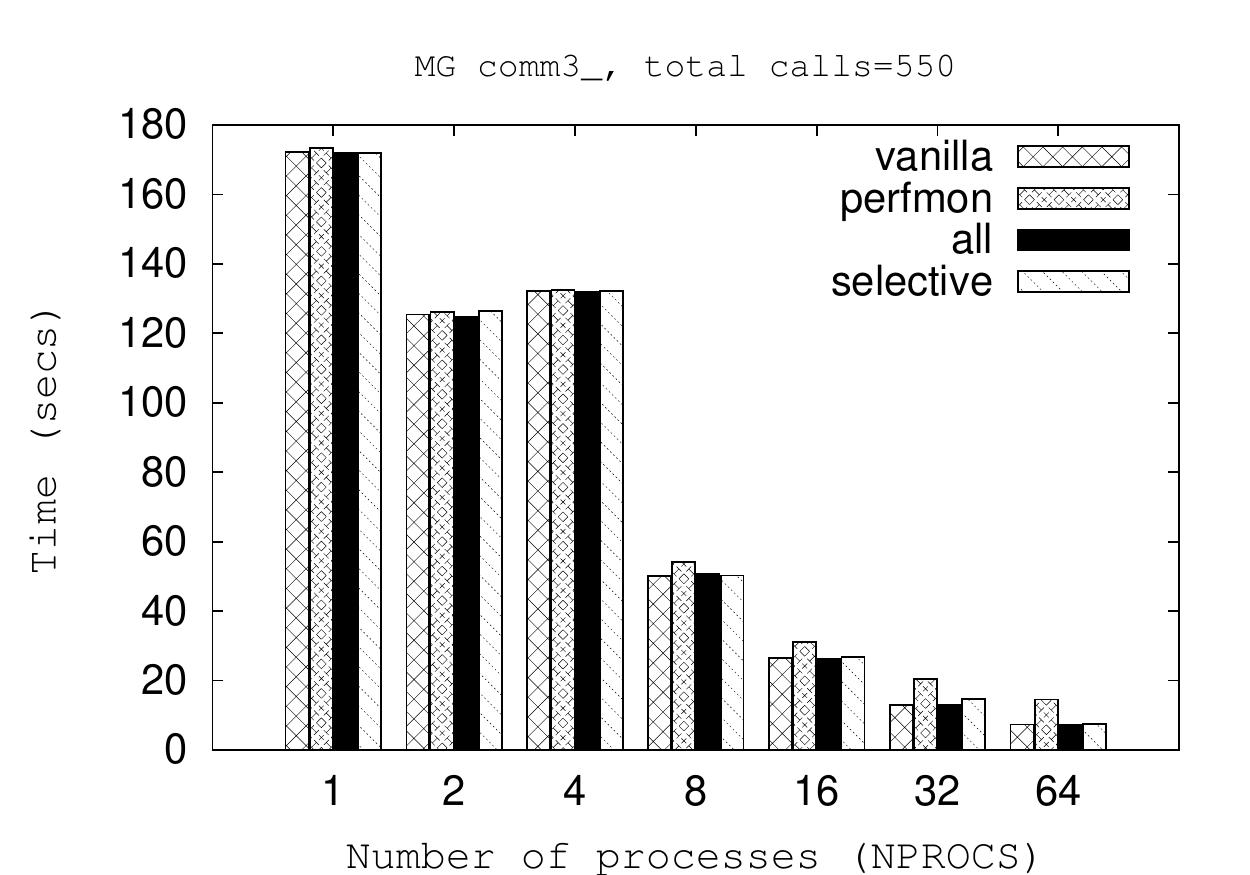}}
\subfloat[SP]{\label{fig:sp1}\includegraphics[scale=0.45]{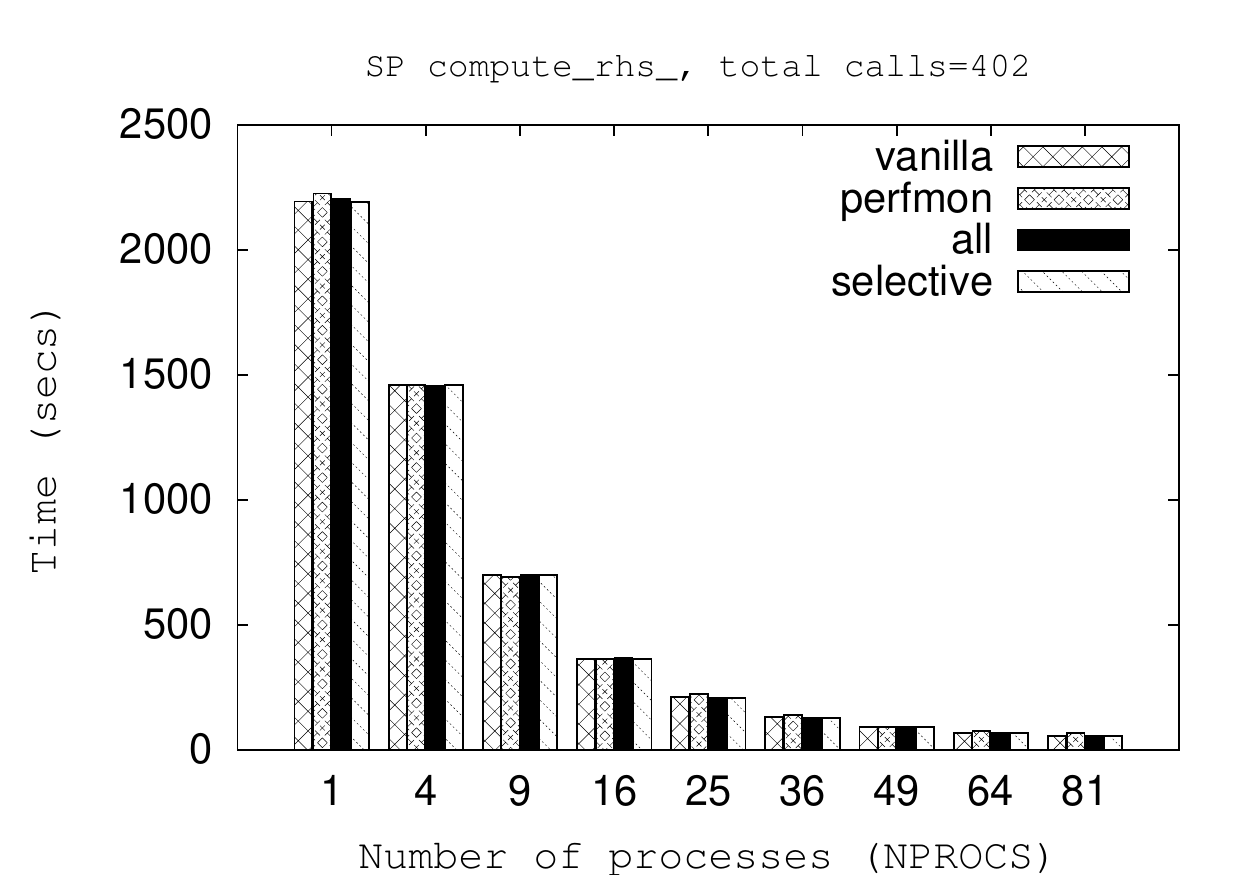}}\\


\caption{Illustrates the performance overhead of various test cases \emph{vanilla}, \emph{perfmon}, \emph{all} and \emph{selective} along with the selected function and its total number of calls.}
\label{figure3}
\end{figure*}

We monitored several functions, one at a time per each benchmark. The choice of these functions is based on the number of times they are called during the entire program execution. In Figure~\ref{figure2} we present results for the functions that were called the maximum number of times. Figure~\ref{figure3} illustrates the results for functions which were called on the order of tens of times, hundreds of times and several thousands of times. We choose the NAS CLASS C workload and ran each benchmark with increasing number of processors $(1<= NPROCS<= 81)$. Finally, we ran each benchmark using \emph{perfmon} (pfmon-3.6). We compiled the sources of Perfmon with debugging support disabled (CONFIG\_PFMON\_DEBUG=n) in the config.mk file and we also commented a \emph{printf} statement that prints ``unknown ptrace event" in Line 2045 in \emph{pfmon\_task.c}. 



We present our findings briefly below:

\begin{itemize}

\item In general, \emph{vanilla} took the least amount of time to execute (shown in Figures~\ref{fig:bt:binvcrhs}--\ref{fig:sp:exact_solution}) compared to the other cases (\emph{perfmon}, \emph{all} and \emph{selective}). This was the expected result.

\item In scenarios where the number of times a particular function is invoked is relatively small (tens of thousands), the performance monitoring overhead of \emph{perfmon} is comparable to \emph{selective} and \emph{all}, as shown in Figure~\ref{fig:mg:randlc}. This is because the overhead of the underlying technique (breakpoints or compiler instrumentation) is independent of the total life time or scope of a particular function; instead, it depends on the number of times the function is called. Hence, in such scenarios, the actual impact of the hardware performance monitoring methodology is insignificant in the total execution time.

\item The overhead of \emph{all} was higher compared to \emph{vanilla} and \emph{selective} for some benchmarks. Since the underlying performance monitoring technique of \emph{all} and \emph{selective} are the same, this overhead is solely attributed to additional cost of intercepting all functions. The extent of this overhead depends on the total number of functions and the number of times each is invoked. We found that in some cases with relatively small number of function calls, (shown in Figures~\ref{fig:is1}-\ref{fig:is2}) \emph{all} had slightly more overhead compared to \emph{Perfmon}.



\item In general, for a majority of benchmarks the overhead of \emph{perfmon} was much higher than ScALPEL; in some cases (Figures~\ref{fig:bt:binvcrhs}-\ref{fig:is:randlc}) by two to three orders of magnitude. We found that the \emph{selective} test case had significantly lower overhead compared to \emph{perfmon}. This overhead varies with function call counts in different benchmarks as shown in Figures~\ref{fig:bt:binvcrhs}-\ref{fig:sp:exact_solution}. The key reason for the reduction in overhead was our use of function interception instead of breakpoints in \emph{perfmon}, which incurs increasing overhead as the monitored function is called repeatedly. Also, the overhead of using \emph{perfmon} to monitor a particular function was higher than the overhead of cumulatively profiling all the functions and monitoring the counters for a given function using \emph{all}. Figure~\ref{figure2} clearly shows that the compiler directed approach provided by ScALPEL provides a low overhead approach to parallel performance measurement.

\end{itemize}

\subsection{Case Study}
In this section we use a simple
case study to illustrate the benefits of runtime reconfigurability as provided by ScALPEL.  Two features of ScALPEL are particularly important in this
case study, namely the ability to profile only specified functions and the ability to
dynamically switch between any number of hardware counters.
Existing tools, such as Perfmon, allow users to change
the counters of interest at regular specified time intervals.  However, switching events at fixed 
time intervals does not yield accurate performance results for function-level profiling 
since there may not be any correlation between the time intervals and the function
call pattern of the application.
In this case study we use ScALPEL to cycle through the event sets of
interest after a fixed number of calls to the function(s) of interest. 
Furthermore, the performance of real applications often depends
subtly on interactions among several factors, e.g., instruction level scheduling
and parallelism, register pressure, memory hierarchy issues, etc.  
Hence, it is extremely difficult to
know, a priori, which few event counters to track in a given performance
gathering run.  With ScALPEL, we can gather data from an arbitrary number of 
performance counters, by cycling through a large set of counters in a predictable way.

As a case study, we use the LINPACK
benchmark to compare the performance of two different implementations of the
Basic Linear Algebra Subprograms (BLAS), viz.
ATLAS{\footnotesize~\cite{Whaley1998}} and
GotoBLAS{\footnotesize~\cite{Goto2008}}. The goal here is not 
to benchmark a system, but rather to study the performance of two implementations 
of the dominant linear algebra kernel underlying that benchmark code. 
We do this by analyzing the hardware counters and not the implementations themselves.

High-performance scientific computations depend to a large degree on the performance 
of the matrix multiplication kernel. The General
Matrix Multiply (GEMM) subroutine in BLAS performs matrix multiplications
$C \leftarrow \alpha AB + \beta C$; $A$, $B$ and $C$ being
matrices, while $\alpha$ and $\beta$ are scalar coefficients. GEMM is often
highly tuned to run as fast as possible for high-performance computing as it is
the building block for many other routines. GEMM is often used recursively,
with input matrices decomposed into smaller block matrices which are in turn
operated on using GEMM. 
Such matrix decompositions allow for better
locality of reference, thereby yielding better utilization of 
system cache. When there is more than one level of cache,
blocking can be applied at each level. This technique is one of the
optimizations used in the implementation of ATLAS, and based on published
reports, one expects the ATLAS implementation of GEMM to perform well
in terms of its use of the memory hierarchy. In GotoBLAS, on the other hand, 
the focus is on minimizing Translation Look-aside Buffer (TLB) table misses. 
According to{\footnotesize~\cite{Goto2002}}, ``While the importance of cache is
also taken into consideration, it is the minimization of such TLB misses that
drives the approach.'' TLB misses are minimized by filling most of the memory
addressable
by the TLB with the matrix $A$, while operating on matrices $C$ and $B$ a few
columns at a time. 


We used ATLAS developmental version 3.9.5 and GotoBLAS version 1.26 in our case study. Both
ATLAS and GotoBLAS were compiled to use only a single thread. ATLAS was
compiled with both the architectural defaults and full search 
(\emph{-Si archdef 0}). Five different sets of events were monitored in a single sampling run of the benchmark
viz. \emph{\{DTLB\_MISSES:ANY and L2\_LINES\_IN:ANY\}}, \emph{\{L2\_RQSTS:ANY
and SSE\_PRE\_MISS:NTA:L1:L2\}}, \emph{\{L1D\_ALL\_REF and
L1D\_ALL\_CACHE\_REF\}}, \emph{\{X87\_OPS\_RETIRED:ANY and
SIMD\_INST\_RETIRED:ANY\}} and \emph{\{INST\_RETIRED:ANY\_P and
RESOURCE\_STALLS:ANY\}} (all in perfmon2 format).
The \emph{ATL\_dgemm} function was instrumented in the
ATLAS implementation, while the \emph{dgemm\_} function was instrumented for
GotoBLAS. We cycled through the event sets of interest after every $100$ calls
to the DGEMM implementation during the sampling run. We also ran the exhaustive case in which we 
monitored one set of events per run and ran the benchmark five times capturing a different set
of events on each run.

\begin{table}[ht]
\centering
\small
\caption{Hardware counter values for LINPACK run NB = 200, N = 20000 sampling
run}
\begin{tabular}{|m{1.3in}|m{0.45in}|m{0.45in}|m{0.45in}|} 
\hline
Event name & ATLAS (default) & ATLAS (full) & Goto \\
\hline
DTLB\_MISSES &  2.78e07 & 2.88e07 & 4.61e07 \\
L2\_LINES\_IN & 1.65e09 & 1.56e09 & 5.72e08 \\
L1D\_ALL\_REF & 2.26e11 & 2.25e11 & 1.52e11 \\
L1D\_ALL\_CACHE\_REF & 2.26e11 & 2.25e11 & 1.52e11 \\
X87\_OPS\_RETIRED & 7.16e05 & 2.66e05 & 0.00e00 \\
SIMD\_INST\_RETIRED & 7.13e11 & 7.13e11 & 7.11e11 \\
INST\_RETIRED & 8.19e11 & 8.19e11 & 8.76e11 \\
RESOURCE\_STALLS & 6.39e10 & 6.35e10 & 1.57e10 \\
\hline
\end{tabular}
\label{tab_counters}
\end{table}

\begin{figure}
\centering
\includegraphics[scale=0.65]{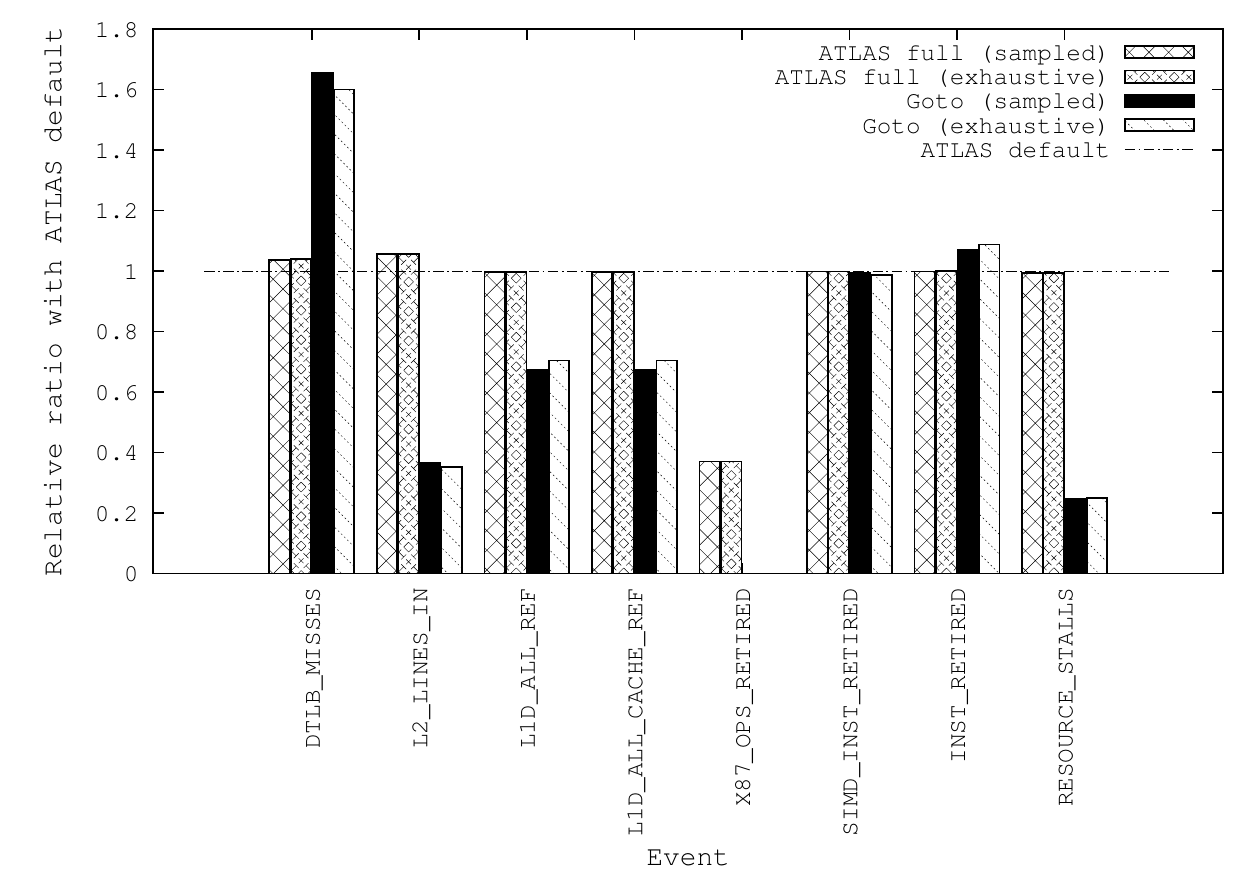}
\caption{Comparison of ATLAS full build and GotoBLAS.}
\label{fig_blas_comparison}
\end{figure}

The raw hardware counter values are presented in
Table~\ref{tab_counters} for problem size $N = 20000$ and block size
$NB = 200$ with sampling. We do not present the results for the second set of
events \emph{\{L2\_RQSTS:ANY and SSE\_PRE\_MISS:NTA:L1:L2\}} as the counters
returned zeros. In Figure~\ref{fig_blas_comparison} we compare the
relative ratio of ATLAS full build and GotoBLAS with the
architectural defaults build of ATLAS for both exhaustive and sampling runs.
Comparing the exhaustive runs to our function call multiplexed sampling 
technique shows that the error introduced by sampling is marginal. 
In terms of bottom-line performance, the benchmark built with GotoBLAS is
$9.5\%$ faster than with the default build of ATLAS, while with ATLAS full
build we get only an $0.6\%$ improvement over the default build of ATLAS.
Normally, all we can do is attribute these performance differences to the cleverness of one
implementation over another; but with ScALPEL we can look closer and understand
in detail why these observed differences occur, and perhaps
identify opportunities for further improvements in this or other codes.

Recall that the stated goal of
the GotoBLAS implementation is to reduce TLB misses. 
Surprisingly, we see from Figure~\ref{fig_blas_comparison}
that GotoBLAS has $65\%$ {\em more\/} TLB misses than ATLAS.
Looking at other counters, however, we see that GotoBLAS 
has $65\%$ fewer L2 cache misses than ATLAS, 
and $75\%$ fewer resource stalls.
It seems clear that the GotoBLAS implementation is gaining a performance advantage from
these significant reductions in expensive events.  Simply counting the total number of 
TLB misses is misleading.  In fact, the Goto implementation incurs substantially
more total TLB misses.  However, it appears that Goto is able to amortize
(or even completely hide) the cost of these misses over a relatively greater amount of useful
computation.  Said another way, not all TLB misses are created equally.  Some may be essentially
harmless, and if the TLB misses are managed and scheduled carefully (e.g., with aggressive
pre-fetching), then an implementation may be able to reduce other expensive events
(e.g., stalls and cache misses), as happens in this case.


\section {Limitations}\label{section_limitations}
In this section we discuss some of the limitations of ScALPEL. First, as discussed previously, while our approach does not require any source code modifications, it does require recompilation. Second, we presently support profiling functions that are recursive and have nested calls to other functions. In such situations we monitor the hardware counters for both the function (parent) and its nested call sequence (children). However, we do not support monitoring both the parent and the child at the same time in the same nested call sequence. Third, the granularity of our profiling is restricted to a function level; we do not support profiling at block level. Finally, our present implementation does not report the results on a per thread basis within a process. This is also an issue with Perfmon. In the future we plan to enhance our prototype to isolate the hardware counter values at the granularity of a thread in multi-threaded application. 

\section {Conclusion}\label{section_conclusion}

In this paper we addressed several shortcomings of existing performance monitoring tools. We demonstrated that it is possible to monitor the hardware counters using ScALPEL without imposing any significant overhead by using a compiler directed instrumentation technique. Moreover, by coupling the instrumentation technique with our runtime system, we were able to provide an efficient performance monitoring scheme.

We provided a prototype to adaptively configure both functions and events at runtime. We discussed the performance implications of our approach and validated its usage with a case study. We discussed the key issues that lead to performance overhead. Our results indicate that our function instrumentation technique outperforms if not matches other existing tools for a majority of benchmarks. Moreover, for certain benchmarks, both our best and worst case scenarios performed significantly better than Perfmon. And most importantly, we accomplish this without the need for any modifications to an application's source code. Finally, our approach is completely portable and can be used with existing implementations such as Perfmon and PAPI.

\bibliographystyle{IEEEtran}
\bibliography{icpp_2009_paper}

\end{document}